\documentstyle[aps,twocolumn,graphics,epsf,amsmath,amssymb,eqsecnum,floats]{revtex}
\begin{document}
\draft
\twocolumn[\hsize\textwidth\columnwidth\hsize\csname @twocolumnfalse\endcsname
\title{Quantum disorder in the two-dimensional pyrochlore Heisenberg
antiferromagnet} 
\author{S. E. Palmer$^{1,2}$ and J. T. Chalker$^1$}
\address{$^1$Theoretical Physics, University of Oxford, 1 Keble Road,
Oxford OX1 3NP, UK} 
\address{$^2$Current address: Sloan Center for
Theoretical Neuroscience, University of California at San Francisco,
Box 0444, 513~Parnassus Avenue, San Francisco, CA 94143, USA}
\maketitle 
\date{\today} 
\maketitle
\begin{abstract}
We present the results of an exact diagonalization study of the
spin-$1/2$ Heisenberg antiferromagnet on a two-dimensional version of
the pyrochlore lattice, also known as the square lattice with
crossings or the checkerboard lattice.  Examining the low energy
spectra for systems of up to 24 spins, we find that all clusters
studied have non-degenerate ground states with total spin zero, and
big energy gaps to states with higher total spin.  We also find a
large number of non-magnetic excitations at energies within this spin
gap.  Spin-spin and spin-Peierls correlation functions appear to be
short-ranged, and we suggest that the ground state is magnetically
disordered.
\end{abstract}
\vskip 0.2 truein
\newpage
\vskip2pc]

\section{Introduction}
Geometrically frustrated magnets have attracted the interest of both
experimentalists and theorists in recent years \cite{reviews},
particularly because of their unusual low-temperature properties.
Fluctuations enhanced by frustration can destroy long range order,
giving rise to disordered ground states even in systems with two or
three space dimensions.  Present understanding of such states remains
incomplete.  In this context it is useful to study a variety of
models, so as to establish the types of behavior that are possible and
to isolate good instances of each type.

Examples of geometrically frustrated lattices include the triangular
and kagom\'e lattices in two dimensions, and the pyrochlore lattice in
three dimensions. All these are based on frustrated units -- triangles
in two dimensions and tetrahedra in three dimensions -- and in the
last two examples these units are combined in a site-sharing
arrangement.  Properties of the spin-$1/2$ Heisenberg antiferromagnet
with nearest-neighbor interactions have been studied for each of these
lattices.  In the case of the triangular lattice, the three sublattice
N\'eel order of the classical ground state is believed to be stable
against quantum fluctuations \cite{bernu94,captrum9900}.  By contrast,
in the case of the spin-$1/2$ kagom\'e Heisenberg antiferromagnet
(KHAF), analysis of the low-energy spectra for finite clusters
indicates a disordered ground state
\cite{zeng90,chalker92,leung93,lecheminant97,waldtmann98}. Calculations
for the three-dimensional spin-$1/2$ pyrochlore Heisenberg
antiferromagnet present a great challenge; an approximate approach
\cite{canals98} yields a disordered ground state in that case, too.
 
\begin{figure}
\begin{center}
\leavevmode
\hbox{%
\epsfysize=2.0in
\epsffile{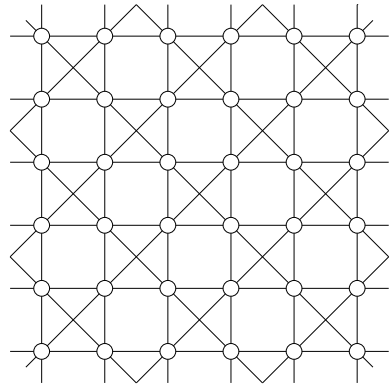}}
\caption{The two-dimensional pyrochlore lattice, or square lattice
with crossings. Circles indicate spins of the Heisenberg model, and
lines denote exchange interactions, all of equal strength.}
\label{slwc}
\end{center}
\end{figure}

In this paper, we describe properties of the spin-$1/2$ Heisenberg
antiferromagnet on a two-dimensional version of the pyrochlore
lattice, also known as the square lattice with crossings or the
checkerboard lattice.  We denote this model by SLWC. The lattice is
illustrated in Fig.\,\ref{slwc}.  The frustrated plaquettes (those
with bond-crossings) have the connectivity of tetrahedra, and are
corner-sharing, as in the three-dimensional pyrochlore lattice. The
classical Heisenberg model on this two-dimensional lattice is known to
have disordered ground states \cite{moessner98b}, while for the
quantum Heisenberg model with general spin $S$ it has been proved
\cite{lieb99} that all ground states have total spin zero.  From
calculations presented below of the low-energy spectra for clusters of
$N$ sites, with $N$ taking even values from $N=10$ to $N=24$, we
conclude that the ground state is disordered at $S=1/2$.

The SLWC joins a rather short list of two-dimensional Heisenberg spin
systems known to have disordered quantum ground states.  Other
examples, apart from the KHAF, involve frustration which is not
geometric but arises either from further neighbor interactions or from
multiple spin exchange.

The effects of further-neighbor interactions have been studied
extensively for the Heisenberg model on the square lattice. With
nearest-neighbor exchange $J_1$ and second neighbor exchange $J_2$,
N\'eel order is destabilized by large ground state degeneracy in the
classical model at an exchange ratio of $J_1/J_2=2$ \cite{chandra88}.
With spin-$1/2$, it is likely that the ground state is magnetically
disordered and has a gap for all excitations
\cite{rokhsar89,dagotto89,gelfand89,singh90,schultz92,schultz96,capriotti00,kotov00,jongh00,sushkov00}. The
SLWC and the $J_1-J_2$ model at $J_1/J_2=2$ are quite closely related,
in the sense that the $J_1-J_2$ model has second-neighbor exchange
interactions which, compared to those in the SLWC, are twice as
numerous but of half the strength.  It has been noted very recently
\cite{canals01} that a linear spin-wave analysis suggests a disordered
ground state for the SLWC, as in the $J_1-J_2$ model.

The consequences of frustration by multiple spin exchange have been
examined for the spin-$1/2$ Heisenberg antiferromagnet on the
triangular lattice \cite{misprl98,misprb98,misprb00}. Competition
between two-spin and four-spin exchange leads to two disordered spin
states: one has an energy gap for excitations, while the other is
similar to the KHAF in having many low-energy excitations.

\section{Model and Methods}
The Hamiltonian we study is
\begin{equation}
{\mathcal H}=2\sum_{\langle i,j\rangle}{\mathbf S}_i\cdot{\mathbf S}_j,
\end{equation}
where the sum is taken over pairs of spins $i$ and $j$ joined by bonds
on the lattice of Fig.\,\ref{slwc} and the ${\mathbf S}_i$ are
spin-$1/2$ operators.  The magnitude of the exchange interaction in
this Hamiltonian is the same as that used for the KHAF in
Ref.\! \cite{lecheminant97}, and is a factor of two larger than that
used in References \cite{zeng90} and \cite{leung93}.

We may also express the Hamiltonian as a sum over tetrahedra, 
equivalent to square plaquettes with crossings, by writing
\begin{equation}
{\mathcal H}=\sum_\alpha\left[\left(\mathbf{J}_\alpha\right)^2-3\right],
\label{H2}
\end{equation}
where $\alpha$ labels the frustrated plaquettes and ${\mathbf
J}_\alpha ={\mathbf S}_a + {\mathbf S}_b +{\mathbf S}_c +{\mathbf
S}_d$ is the total spin of plaquette $\alpha$, to which the sites $a$,
$b$, $c$ and $d$ belong.  Expressing the eigenvalues of $|{\mathbf
J}_\alpha|^2$ as $J(J+1)$, the quantum number $J$ takes the values 0,1
or 2.  Since each spin is a member of two plaquettes with crossings,
the operators ${\mathbf J}_\alpha$ are not all independent. In
consequence, it is not possible to construct a state in which $J=0$
for every $\alpha$.  Nevertheless, we show below that $J=0$ dominates
and that $J=2$ has very small weight.

We use symmetry under spin rotations to block-diagonalize the
Hamiltonian, and employ the Lanczos algorithm to calculate the
low-lying eigenvalues for clusters of up to 24 spins. We find
eigenvectors by inverse iteration.  We test the accuracy of an
eigenvalue obtained after $j$ steps of the Lanczos algorithm by
performing an additional $j/10$ steps and then recalculating the
eigenvalue: we report only results for which the eigenvalue estimate
changes by less than $10^{-6}$ under this test.

As a check of the code, we study 12 and 21 site clusters for the
spin-$1/2$ triangular lattice Heisenberg antiferromagnet and the KHAF,
examined previously in References \cite{leungrun93} and \cite{zeng90}.
We reproduce the energy levels and degeneracies to the accuracy
reported in each of those papers.

\begin{figure}
\begin{center}
\leavevmode
\hbox{%
\epsfxsize=3.25in
\epsffile{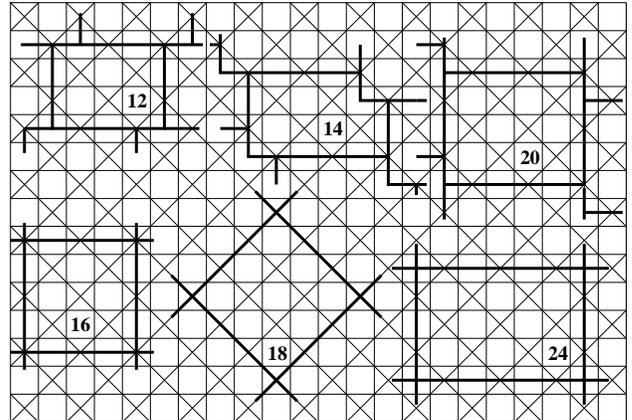}}
\caption{The cluster shapes used in the exact diagonalization
calculations.}
\label{clusters}
\end{center}
\end{figure}

The clusters we study for the SLWC are shown in
Fig.\,\ref{clusters}. For some cluster sizes, several inequivalent
shapes are possible.  In these cases we choose the shape for which the
ground state energy is lowest, on the basis that these shapes
presumably have the least frustrating boundary conditions. In
particular (with the exception of the 14-site cluster) the clusters of
Fig.\,\ref{clusters} leave unfrustrated the two antiferromagnetically
ordered states found in the $J_1$-$J_2$ model, at small and large
$J_2/J_1$ respectively \cite{chandra88}.

\section{Results}

\subsection{Energy level spectra}
\begin{figure}
\begin{center}
\leavevmode
\hbox{%
\epsfxsize=3.25in
\epsfysize=1.7in
\epsffile{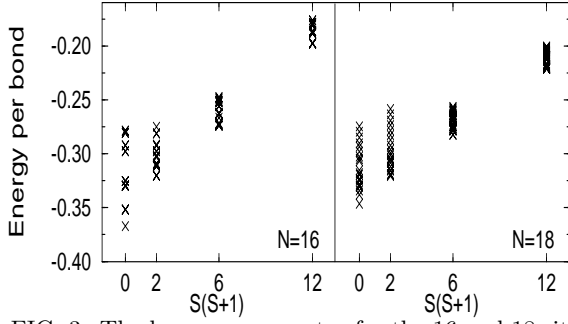}}
\caption{The low energy spectra for the 16 and 18 site clusters for
the SLWC.}
\label{spectra1618}
\end{center}
\end{figure}

\begin{figure}
\begin{center}
\leavevmode
\hbox{%
\epsfxsize=3.25in
\epsfysize=1.7in
\epsffile{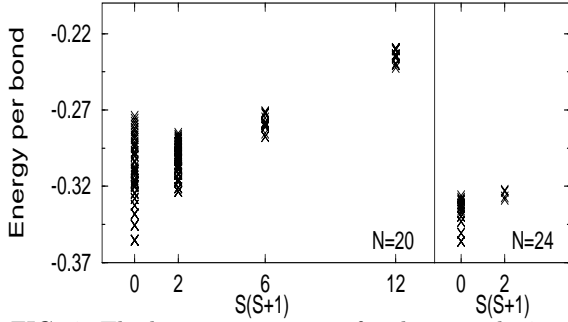}}
\caption{The low energy spectra for the 20 and 24 site clusters for the SLWC.}
\label{spectra2024}
\end{center}
\end{figure}

The low-energy spectra for the 16, 18, 20, and 24 site clusters are
displayed in Figures \ref{spectra1618} and \ref{spectra2024}, labeled
by the total spin quantum number, $S$. They are also listed in Table
\ref{16182024valdeg}. We notice several distinctive features of these
results.  First, the ground state for each cluster is non-degenerate,
and has total spin zero as expected from the rigorous results of
Ref.\,\cite{lieb99}.  Second, there is no evidence in the spectra of
the structure which one would expect if the system had N\'eel order in
the thermodynamic limit.  More specifically, for a N\'eel-ordered
system the quantum mechanics of the order parameter gives rise to a
set of low-lying energy levels
\cite{anderson52,fisher89,gross89a,gross89b,neuberger89,tang89},
termed quasi-degenerate joint states in recent studies
\cite{bernu94,lecheminant97}. These are separated in energy from other
states and have an excitation energy that varies with the total spin
quantum number $S$ and the cluster size $N$ as $S(S+1)/N$. States of
this kind are clearly identifiable in the spectra of triangular
lattice Heisenberg antiferromagnet clusters from the same size range
\cite{bernu94}.  Their absence in the SLWC suggests that the ground
state of the SLWC does not possess N\'eel order.  Third, as we discuss
below in more detail, the number of states lying within the triplet
spin gap is large and increases with cluster size.  Interestingly,
both the absence of quasi-degenerate states and the presence of
low-lying singlet excitations are also features of the KHAF
\cite{lecheminant97,waldtmann98}.

We comment in passing that the data of Table \ref{16182024valdeg}
include some initially unexpected degeneracies for excited states,
involving factors of 3 and 5. A 3-fold degeneracy on the 16-site
cluster can be understood by recalling that each crossed square has
the same connectivity as a tetrahedron, which has four axes of 3-fold
rotational symmetry. Analogously, cycles of length 3 arise from
permutations of the tetrahedra in the 16 site cluster.  We believe
that other 3-fold degeneracies appear in a similar way, and that
5-fold degeneracies on the 20-site cluster are associated with
translations.

\begin{table}
\begin{center}
\caption{Lowest eigenvalues given as an average energy per bond, total
spin, $S$, and degeneracies for various sample sizes on the square
lattice with crossings.  A `-' indicates that the degeneracies of
these levels have not been calculated.  Degeneracies do not include
the $2S+1$ $m_S$ degeneracy.}
\label{16182024valdeg}
\begin{tabular}{cccc}
$N$&
$\overline{2\langle{\mathbf S}_i\cdot{\mathbf S}_j\rangle}$&$S$&deg.\\
\hline 
16&-0.367045&0&1\\ 
&-0.351695&0&4\\
&-0.329769&0&4\\
&-0.329403&0&12\\
&-0.325200&0&6\\
&-0.320232&1&9\\
\hline
18&-0.346423&0&1\\
&-0.337917&0&4\\
&-0.334194&0&1\\
&-0.330624&0&4\\
&-0.329751&0&1\\
&-0.328540&0&1\\
&-0.324801&0&8\\
&-0.323410&0&4\\
&-0.320797&1&-\\
\hline
20&-0.359897&0&1\\
&-0.350021&0&2\\
&-0.342692&0&2\\
&-0.336797&0&5\\
&-0.332620&0&5\\
&-0.329264&0&5\\
&-0.328645&0&5\\
&-0.327938&1&-\\
\hline
24&-0.360776&0&1\\
  &-0.355258&0&1 \\
  &-0.347472&0&2 \\
  &-0.344195&0&1 \\
  &-0.344142&0&1 \\
  &-0.341941&0&6 \\
  &-0.340329&0&1 \\
  &-0.338475&0&2 \\
  &-0.337994&0&6 \\
  &-0.337811&0&6 \\
  &-0.336385&0&1 \\
  &-0.335848&0&2 \\
  &-0.334265&0&6 \\
  &-0.333261&0&3 \\
  &-0.332998&1&- \\
\end{tabular}
\end{center}
\end{table}
\subsection{Dependence on cluster size}
\label{scaling}
\begin{figure}
\begin{center}
\leavevmode
\hbox{%
\epsfxsize=3.25in
\epsffile{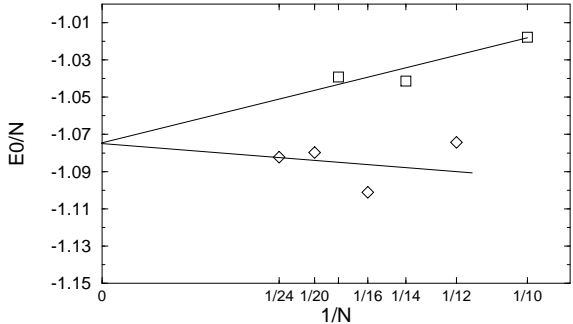}}
\caption{The ground state energy per spin plotted versus $1/N$, where
$N$ is the number of spins the system.  The two lines shown are guides
to the eye and are drawn through the data for the larger cluster
sizes, with an even (diamonds) and odd (squares) number of unit cells
respectively.}
\label{E0scaling}
\end{center}
\end{figure}

The variation with cluster size of the ground state energy per spin is
illustrated in Fig.\,\ref{E0scaling}.  The unit cell of the lattice
contains two sites, and so it is natural to group clusters with even
and odd numbers of unit cells separately. Doing this, a common
extrapolation to the infinite system can be found, as shown.  The
ground state energy as well as the triplet and singlet gaps are
plotted versus $1/N$. (This choice is made simply as a convenient way
of separating points along the horizontal axis.) The cluster sizes
studied here are small, and there are clear difficulties in drawing
quantitative conclusions from the data.  Nevertheless, a definite
indication emerges that the ground state is magnetically disordered.

\begin{figure}
\begin{center}
\leavevmode
\hbox{%
\epsfxsize=3.25in
\epsffile{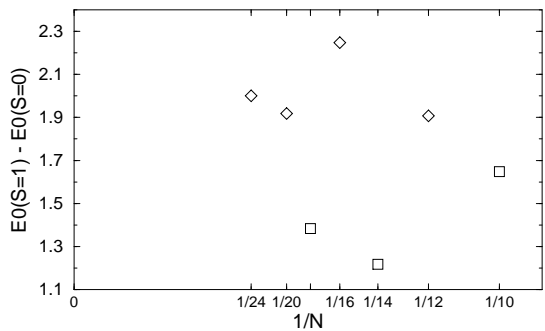}}
\caption{The gap to the first excited state with higher total spin
plotted versus inverse system size.}
\label{spingapscaling}
\end{center}
\end{figure}

The dependence of the spin gap on cluster size is presented in
Fig.\,\ref{spingapscaling}. The fact that the spin gap appears to
remain non-zero in the thermodynamic limit is, of course, in contrast
to the size-dependence of the longest wavelength magnon energy in a
system with N\'eel order.  It suggests again that the SLWC does not
have N\'eel order.  A non-zero spin gap is also believed to be a
feature of the KHAF, but the spin gap for the KHAF is substantially
smaller than that measured for the SLWC.  For example, for the 21 site
cluster on the kagom\'e lattice, Leung and Elser report a spin gap of
size 0.5573\cite{leung93} (scaled to accommodate our factor of 2 in
the Hamiltonian), compared to a spin gap of size $2.00$ for our 24
site cluster. Similarly, the $J_1-J_2$ model at $J_1/J_2=2$ has a
non-zero spin gap, but one that is smaller by a factor of about 2 at
$N=24$ than for the SLWC \cite{capriotti00}.

\begin{figure}
\begin{center}
\leavevmode
\hbox{%
\epsfxsize=3.25in
\epsffile{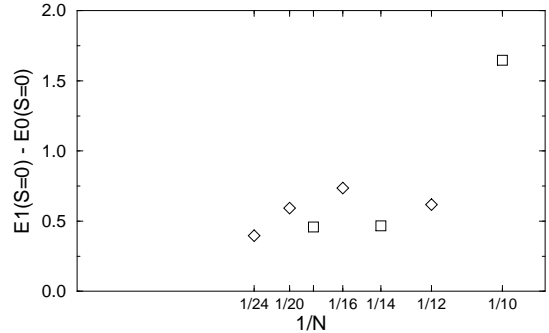}}
\caption{The gap to the first excited state above the ground state is
plotted versus inverse system size.  For all clusters, the first
excited state is also a singlet state.}
\label{singletgap}
\end{center}
\end{figure}

\subsection{Moments of ${\mathbf J}_\alpha^2$}
The fact that the Hamiltonian can be expressed as in Eq.\,\ref{H2}, in
terms of the total spins ${\mathbf J}_\alpha^2$ of the plaquettes with
crossings, leads naturally to the question of what component the
ground state wavefunction has in sectors with each of the possible
values for the quantum number $J$ for a given ${\mathbf J}_\alpha^2$.
Expanding the ground state wavefunction $|\psi\rangle$ as
\begin{equation}
|\psi\rangle = a_0|0\rangle + a_1|1\rangle +
a_2|2\rangle,
\end{equation}
where ${\mathbf J}_\alpha^2|J\rangle=J(J+1)|J\rangle$ for $J=0,1$ or
$2$, we determine the coefficients $a_J$ by measuring
$\langle\psi|{\mathbf J}_\alpha^2|\psi\rangle$ and
$\langle\psi|{\mathbf J}_\alpha^4|\psi\rangle$.  Performing this
calculation on the ground state for the 24 site cluster, we obtain
$\langle {\mathbf J}_\alpha^2\rangle =0.83534$ and $\langle {\mathbf
J}^4_\alpha\rangle =2.01968$ for every plaquette $\alpha$.  This
yields $|a_0|^2=0.611$, $|a_1|^2=0.374$, and $|a_2|^2=0.015$. For
comparison, a covering of the lattice with nearest-neighbor singlets
yields $|a_0|^2=0.25$, $|a_1|^2=0.75$ and $|a_2|^2=0.0$. Since our
measured value of $|a_2|^2$ is small, we conclude that the ground
state lies almost entirely within the sub-space of states in which
every plaquette with crossing contains a nearest-neighbor singlet. In
addition, it is clear that resonance between different such singlet
coverings must make a central contribution to the ground state energy.
These results may serve as a constraint in future attempts to
construct an approximate dimer description of the ground state, for
instance along the lines of Ref.\, \cite{mambrini00}.

\subsection{Non-magnetic states in the spin gap}
The presence of a large number of non-magnetic excited states in the
spin gap is a feature that the SLWC shares with the KHAF.  These
states are spread in energy across the spin gap, and the lowest of
them defines the smallest excitation energy in the system: the singlet
gap.  The dependence of the singlet gap on cluster size is shown in
Fig.\,\ref{singletgap}. The gap is broadly decreasing with increasing
cluster size, but it is unclear from these data whether it is non-zero
or vanishing in the thermodynamic limit. In any case, it is very much
larger than the singlet gap on kagom\'e clusters of similar size: we
find for the SLWC with $N=24$ a singlet gap of $0.397$, to be
contrasted with a singlet gap of $3.02 \times 10^{-3}$ for the $N=27$
kagom\'e cluster \cite{lecheminant97}.

\begin{figure}
\begin{center}
\leavevmode
\hbox{%
\epsfysize=2.0in
\epsffile{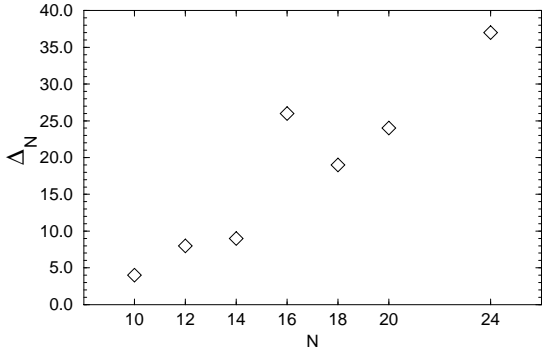}}
\caption{The number of non magnetic ($S=0$) states in the spin
gap plotted versus system size.}
\label{gapstates}
\end{center}
\end{figure}

The total number of singlet states within the spin gap is large and
grows with cluster size.  In Figure \ref{gapstates} we show the number
of states, $\Delta_N$, in the spin gap for each cluster size studied.
The large, anomalous jump in $\Delta_N$ for the 16 site cluster is,
presumably, a finite size effect particular to that sample.  A
comparison can be made with results for the KHAF described in Ref.\,
\cite{lecheminant97}.  At $N=24$, $\Delta_N=37$ for the SLWC and
$\Delta_N=34$ for the KHAF, while more generally for the KHAF
$\Delta_N$ grows with (even) $N$ as $1.14^N$.

\subsection{Ground state correlations}
Finally, we report on the measurements of the ground state correlation
functions for the SLWC.  We have calculated spin-spin and spin Peierls
correlation functions for the 24 site cluster.

As expected both from the absence of quasi-degenerate joint states and
from the large spin gap, the behavior of the spin-spin correlation
function suggests that the SLWC does not have N\'eel order. Values are
give in Table \ref{24sscorr}. Correlations fall off rapidly with
distance and are much smaller in magnitude than in comparison systems
which are believed to have N\'eel order (for example, the spin-$1/2$
Heisenberg antiferromagnet on the triangular lattice, studied in a 36
site cluster in References \cite{bernu94} and \cite{captrum9900}).
Spin-spin correlations on the SLWC are similar in magnitude, but
slightly larger than, the spin-spin correlations on the same sized
cluster for the KHAF \cite{chalker92,leung93}.

\begin{table}
\begin{center}
\caption{Spin-spin correlations for the 24 site cluster. Correlations
are measured between spins at various separations, $r$, from spin 9
(as labeled in Figures 9, 10, and 11).  The nearest-neighbor distance
is taken to be unity.}
\label{24sscorr}
\begin{tabular}{ccccc}
&$n$&$r$&$\langle{\mathbf S}_9\cdot{\mathbf S}_n\rangle$\\
\hline
&15&1&-0.378658\\
&14&$\sqrt{2}$&\phantom{-}0.048430\\
&16&$\sqrt{2}$&-0.081253\\
&21&2&\phantom{-}0.131233\\
&20&$\sqrt{5}$&\phantom{-}0.048430\\
&19&2$\sqrt{2}$&\phantom{-}0.002023\\
&23&2$\sqrt{2}$&\phantom{-}0.002023\\
&12&3&-0.002202\\
\end{tabular}
\end{center}
\end{table}

\begin{figure}
\begin{center}
\leavevmode
\hbox{%
\epsfxsize=3.1in
\epsfysize=2.0in
\epsffile{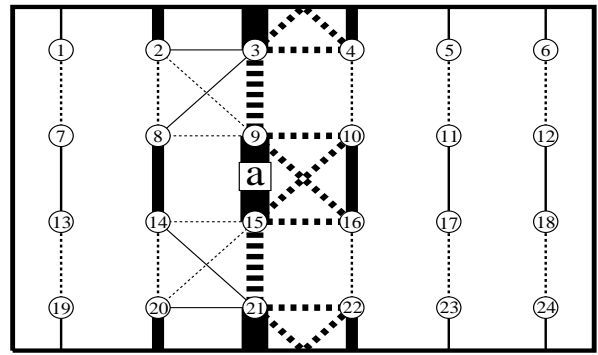}}
\caption{A coarse-grained pictorial representation of the spin Peierls
correlation functions measured for the 24 site cluster relative to the
bond labeled $a$.  Line widths are scaled with the magnitude of the
correlation function.  Positive correlations are given a solid line
while negative correlations are represented with a dashed
line. Correlations that are less than $8\%$ of the on-bond correlation
are omitted.  The values of the correlation functions are given in
Table III.}
\label{24corrfig}
\end{center}
\end{figure}

\begin{figure}
\begin{center}
\leavevmode
\hbox{%
\epsfxsize=3.1in
\epsfysize=2.0in
\epsffile{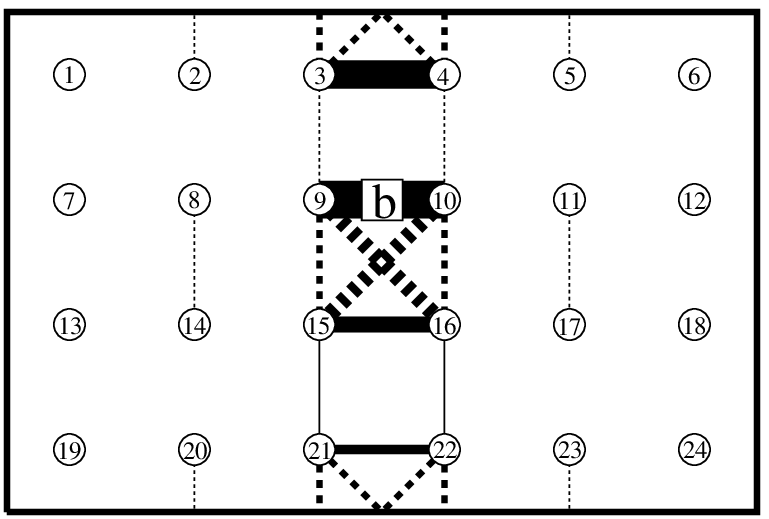}}
\caption{Spin Peierls correlations measured relative to bond $b$ and
represented as described in the caption of Figure 9.}
\label{24corrfigb}
\end{center}
\end{figure}

\begin{figure}
\begin{center}
\leavevmode
\hbox{%
\epsfxsize=3.1in
\epsfysize=2.0in
\epsffile{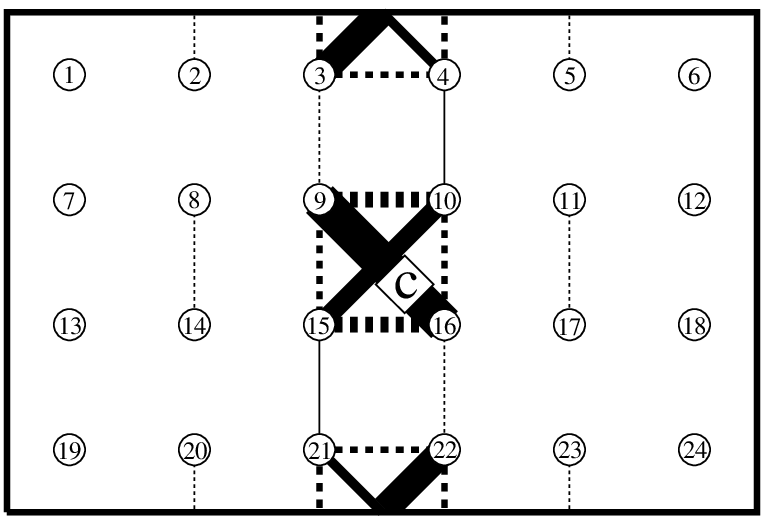}}
\caption{Spin Peierls correlations measured relative to bond $c$ and
represented as described in the caption of Figure 9.}
\label{24corrfigc}
\end{center}
\end{figure}

An alternative to N\'eel order as a pattern for symmetry breaking is
spin Peierls or dimer order. To examine this possibility, we measure
correlations between bond operators $D_a={\mathbf S}_i\cdot{\mathbf
S}_j$, where the bond $a$ joins nearest-neighbor spins $i$ and $j$. We
evaluate the connected correlation function
\begin{equation}
C_P(a,b)=\langle \tfrac{1}{2}(D_aD_b+D_bD_a)\rangle -\langle D_a\rangle
\langle D_b\rangle.
\end{equation}
The results are displayed in Figures \ref{24corrfig}, \ref{24corrfigb}
and \ref{24corrfigc}, and listed in Table \ref{24spcorr}.  The
behavior of this correlation function is clearly influenced by finite
cluster size and periodic boundary conditions. It is striking
nevertheless that the correlation length measured along the long side
of the cluster is no more than one lattice spacing. The magnitudes of
spin Peierls correlations are comparable to those reported for the
KHAF \cite{chalker92,leung93}.

\begin{table}
\begin{center}
\caption{Spin Peierls correlations in the 24 site cluster of the SLWC.
Spin Peierls correlations are measured between the bonds labeled $a$,
$b$, and $c$, and all other bonds, $y$, in the cluster as pictured in
Figures \ref{24corrfig}, \ref{24corrfigb}, and \ref{24corrfigc}.  A
bond $y$ is represented in the table by a pair numbers which label the
two sites joined by $y$.  Site numbers are as given in the
aforementioned figures.}
\label{24spcorr}
\begin{tabular}{cll}
$C_P(a,y)$&bonds $y$\\
\hline
\phantom{-}0.233447&(9,15)\\
\phantom{-}0.193017&(3,21)\\
-0.110574&(3,9),(15,21)\\
\phantom{-}0.083771&(10,16)\\
\phantom{-}0.082317&(4,22)\\
\phantom{-}0.079108&(8,14),(2,20)\\
-0.051080&(9,16),(9,10),(10,15),(15,16)\\
-0.049528&(3,4),(3,22),(4,21),(21,22)\\
\phantom{-}0.034932&(1,19),(5,23),(7,13),(11,17)\\
\phantom{-}0.031582&(6,24),(12,18)\\
-0.026960&(2,8),(4,10),(14,20),(16,22)\\
-0.026735&(5,11),(17,23)\\
-0.020674&(1,7),(13,19)\\
-0.020024&(6,12),(18,24)\\
\phantom{-}0.020004&(2,3),(3,8),(14,21),(20,21)\\
-0.018660&(2,9),(8,9),(14,15),(15,20)\\
\hline
$C_P({\mathrm {\mathit b}\ or\ {\mathit
 c}},y)$&$y$ for $C_P(b,y)$&$y$ for $C_P(c,y)$\\
\hline
\phantom{-}0.221524&(9,10)&(9,16)\\
\phantom{-}0.155176&(3,4)&(3,22)\\
\phantom{-}0.109339&(15,16)&(10,15)\\
-0.101267&(9,16),(10,15)&(9,10),(15,16)\\
\phantom{-}0.076887&(21,22)&(4,21)\\
-0.056178&(3,22),(4,21)&(3,4),(21,22)\\
-0.051080&(9,15),(10,16)&(9,15),(10,16)\\
-0.049528&(3,21),(4,22)&(3,21),(4,22)\\
\phantom{-}0.020004&
(15,21),(16,22)&(4,10),(15,21)\\
-0.018660&(3,9),(4,10)&(3,9),(16,22)\\
-0.017738&(2,20),(5,23),&(2,20),(5,23),\\
&(8,14),(11,17)&(8,14),(11,17)\\
\end{tabular}
\end{center}
\end{table}

\section{Discussion}

In summary, the results described in Sec. III combine to demonstrate
that the SLWC has a ground state with unbroken spin rotation
symmetry. The lack of N\'eel order is signaled by the absence of
quasi-degenerate joint states, by the magnitude and cluster size
dependence of the spin gap, and most directly by the spin-spin
correlation function itself. It is less clear whether or not the
translational symmetry of the model is broken in the ground state,
although the short range of the spin Peierls correlations along one
direction in Figures \ref{24corrfig} - \ref{24corrfigc} is suggestive
of unbroken translational symmetry. If translational symmetry is
broken, one might expect either columnar dimerization or plaquette
resonating valence bond states, as considered for the $J_1-J_2$ model
near $J_1/J_2=2$
\cite{dagotto89,gelfand89,singh90,capriotti00,kotov00}. Since only a
few examples are known of two-dimensional Heisenberg models with
magnetically disordered ground states, it is a problem of some
theoretical interest to decide between these alternatives and,
especially, to understand the nature of the low-lying singlet
excitations.

\section*{acknowledgments}
We are grateful to P. Lecheminant and G. Misguich for very helpful
discussions.  S.E.P. would like to thank the Rhodes Trust and the
Alfred P.\ Sloan Foundation for financial support.  The work was also
supported in part by EPSRC grants GR/J78327 and GR/M56234.  Some
preliminary results for small (up to 12-site) clusters were obtained
using TITPACK v.\! 2 \copyright\ Hidetoshi Nishimori.

\end{document}